\documentclass[twocolumn]{aastex63}
\usepackage[english]{babel}

\received{November 1, 2020}
\revised{}
\accepted{\today}
\submitjournal{ApJL}
\pdfoutput=1
\shorttitle{Kepler-411 differential rotation}
\shortauthors{Araújo \& Valio.}
\usepackage{epstopdf}

\usepackage{float}

\begin{document}

\title{Kepler-411 differential rotation from three transiting planets}

\correspondingauthor{Alexandre Araújo}
\email{adesouza.astro@gmail.com}

\author[0000-0002-2106-4332]{Alexandre Araújo}

\author[0000-0002-1671-8370]{Adriana Valio}

\affiliation{Center for Radio Astronomy and Astrophysics Mackenzie, Mackenzie Presbyterian University\\
Rua da Consolação, 860, São Paulo, SP - Brazil}

\begin{abstract}

The differential rotation of the Sun is a crucial ingredient of the dynamo theory responsible for the generation of its magnetic field.  Currently, the rotation profile of a star that hosts one or more transiting planets can be estimated. By detecting the same spot in a later transit, it is possible to infer the stellar rotation period at that latitude.  In this work,  we apply for the first time transit spot  mapping to determine the differential rotation of Kepler-411, a K2V-type star with an average rotation period of 10.52 days, radius of 0.79 R$_\odot$ and mass of 0.83 M$_\odot$. Kepler-411 hosts at least four planets, the inner planet is a Super-Earth with a radius of 1.88 R$_\oplus$ and an orbital period of 3.0051 days, whereas the two larger transiting planets are mini Neptunes with radii of 3.27 and 3.31 R$_\oplus$, and periods of 7.834435 and 58.0204 days, respectively. Their orbits are such that they transit the star at latitudes of -11$^{\circ}$, -21$^{\circ}$, and -49$^{\circ}$. Analysis of the transit light curves of the three planets resulted in the detection of a total of 198 spots. For each transit latitude, the rotation period of the star was estimated and the differential rotation pattern estimated independently. Then a solar like differential rotation profile  was fit to the three rotation periods at the distinct latitudes, the result agreed extremely well with the previous ones, resulting in  a differential shear of $0.0500\pm0.0006$ rd/d or a relative differential rotation of $8.4\pm0.1$\%.

\end{abstract}

\keywords{star, spot --- star, rotation --- exoplanets, transit}

\section{Introduction}\label{sec:intro}

Studying magnetic activity on stars other than the Sun provides an opportunity for detailed tests of solar dynamo models. Using only solar observations limits the range of the global stellar parameters for such tests, while an extensive sample of stars of various masses, ages, and activity levels provides important constraints for stellar and solar dynamo theory (Berdyugina 2005).  Turbulent dynamos operating in other stars produce strong magnetic fields and are able to transform poloidal into toroidal fields, and vice-versa. 

Differential rotation is believed to be one major ingredient of the driving mechanism of magnetic field generation on the Sun. Solar differential rotation can be measured using surface features such as sunspots observed at different latitudes during consecutive days. Moreover, the solar rotation profile is by no means uniform, presenting differential rotation with the equator rotating faster than the poles.

Helioseismology reveals that the outer convective region also has a spread of rotation rates with different latitudes. Schou et al. (1998) reported  on joint helioseismic analyses of solar rotation in the convection zone and in the outer part of the radiative core.  The data  revealed that the angular velocity is distinctly lower at high latitudes than the values previously extrapolated from measurements at lower latitudes based on surface Doppler observations and helioseismology.

Likewise, a key ingredient for stellar dynamo models is the stellar differential rotation, which is hard to measure because surface features can only be resolved on the Sun. 
Among the techniques to estimate differential rotation of stars are the Fourier transform of spectral line profiles (Reiners \& Schmitt 2003;
Takeda 2020) and Doppler Imaging of stars that are sufficiently fast rotators ($v\sin i \ge$ 15 km s$^{-1}$). For late-type stars that are slowly rotating ($v\sin i \le$ 12--15 km s$^{-1}$) (Lanza et al. 2014), the rotation profile is determined by asteroseismology (Appourchaux
et al. 2008) and studies based on photometric variability from starspots at different latitudes (Korhonen \& Elstner 2011;
Reinhold et al. 2013). 
Differential rotation is considered ``solar-like'' when the equator rotates faster than the poles, and conversely  when the polar regions rotate faster than the equator, it is said to be ``antisolar''. 

For main sequence stars, the rotation rate strongly depends on the stellar age. Because of rotational braking due to a stellar wind, stars lose angular momentum over time and slow down. 
What is known is that stellar rotation coupled with convective motions generates strong magnetic fields in the stellar interior and produces a multitude of magnetic phenomena including starspots in the photosphere, chromospheric plages, coronal loops, and flares (Berdyugina 2005).

Since the discovery of the first indications of starspots, photometry remains the most common technique for studying stellar activity. Thanks to efforts of individual observers, more or less regular observations of spotted stars began in the 1970’s. More recently, during the last decade, high precision photometry by space missions made possible detailed studies of solar-like  phenomena in other stars. Crucial to studies of stellar activity was the Kepler mission(Borucki et al. 2010), which offered almost continuous observations for 150,000 stars over 4 years.

Therefore, detecting starspots located at different latitudes would be useful tracers of the differential rotation. The analysis of planetary transits is able to detect small variations in the light curve caused by the passage of a planet in front of a solar-like spot on the stellar surface (Silva 2003). These spots can produce different effects on the transit shape. For example, spots that are not occulted by the planet will produce a deeper transit, whereas spots occulted by the transit cause an increase in the luminosity flux during a few minutes (Pont et al. 2013).
 
By detecting the same spot on a later transit it is possible to determine the rotation period of the star at the  latitude occulted by the planet. This method has been successfully applied to the G-type stars CoRoT-2(Silva-Valio et al. 2010), Kepler-17
(Valio et al. 2017), Kepler-63 (Netto \& Valio 2020), and
Kepler-71 (Zaleski et al. 2019), as well as to the M-dwarf Kepler-45 (Zaleski et al. 2020). However, this is the first time that the method will be applied to star with three transiting exoplanets.

 To determine the rotation profile, the latitude of a spot needs to be known (Balona \& Abedigamba 2016). 
The monitored study of these starspots at different latitudes allows the determination of the differential rotation profile of the star. 
In a multi-planetary system with orbits aligned close to the stellar equator, each planet will occult different latitudes of the star. Thus, spots may be detected in more than one latitude of the stellar disc.

In this work we apply the model described in Silva (2003), to characterize the starspots identified on the transits of three different planets. By identifying the same spot on a later transit, it is possible to estimate the rotation period of the star at that latitude. 
Combining the stellar rotation period at each planet crossing latitude, and assuming a solar-like differential rotation profile, we determined the differential rotation profile of Kepler-411. In the next section we present the data and the model. Section~\ref{sec:results} describes our main results of spot properties and stellar rotation, which are discussed in the last section along with our main conclusions.

\section{Observation and modeling} \label{sec:style}

In this study, we analyzed the light curve of the Kepler-411 star (KIC 11551692) observed by the Kepler space telescope for about 600 days, exhibiting characteristics that indicate relatively strong magnetic activity (Sun et al. 2019).
 The MAST data archive contains 17 quarters of data for Kepler-411, both long and short cadence (the latter available only for Q11 to Q17). We used short ($\sim 1$min) cadence data in the Pre-search Data Conditioning (PDCSAP) format for our analysis. 
 
Kepler-411 is a star of spectral type K2V, with a mass of $0.83^{+0.04}_{-0.10}$ M$_\odot$ and radius of $0.79^{+0.07}_{-0.06}$ R$_\odot$ (Wang et al. 2014), located at a distance of 153.6 $\pm$ 0.5 pc.
Three planets, one SuperEarth and two miniNeptunes, are known to transit the Kepler-411 star, which main characteristics  are listed in Table~\ref{tab:param}. There is a fourth planet in orbit of Kepler-411, but since it does not transit the star, it is not considered in this study.

\begin{table*}
\setlength{\arrayrulewidth}{2\arrayrulewidth}
\centering
\caption{Physical parameters of Kepler-411 planets, spots, and differential rotation}\label{tab:param}
\def\arraystretch{1.2}  
\begin{tabular}{l|cccc}
\hline 
\hline
\multicolumn{5}{c}{Planetary Parameters}\\
\hline
& Kepler-411b & Kepler-411c & Kepler-411d \\
\hline
Orbital Period [days]$^{a}$   & $3.0051 \pm 0.00005$ & $7.834435 \pm 0.000002$ & $58.02 \pm 0.0002$ & \\
Planet Radius [R$_\oplus$]$^b$ & $1.88 \pm 0.02$ & $3.27^{+0.011}_{-0.006}$ & $3.31 \pm 0.009$ & \\
Planet Radius [$R_{star}$]$^c$ &   $0.024 \pm 0.002$ &  $0.042  \pm 0.002$ &  $0.040 \pm 0.002$ \\
Semi-Major Axis [au]$^c$  & $0.049  \pm 0.0006$ &
  $0.080 \pm 0.001$   &  $0.29 \pm 0.0004$ & \\
Orbital inclination  $^c$       & $89.18 \pm 0.05$ & 
     $89.03 \pm 0.02$ & $89.44 \pm 0.01$ & \\
\hline
\multicolumn{5}{c}{Spot average parameters}\\
\hline
Transits of & Kepler-411b & Kepler-411c & Kepler-411d & all \\
 \hline
Number of spots & 45 & 143 & 10 & 198 \\
Temperature [K] &  $3100\pm 800$ & $4100\pm 500$ & $3600\pm 800$ & \\
Radius [$10^3$ km] & $18\pm 7$ & 
$17\pm 8$ & $17\pm 5$  & \\
Intensity [$I_c$] & $0.14\pm 0.23$ & $0.46\pm 0.28$ & $0.29\pm 0.19$ & \\
\hline
\multicolumn{5}{c}{Rotation}\\
\hline
Transit latitude [$^\circ$] & $-11 \pm 3$  & $-21.6 \pm 2.3$ & $-49 \pm 7$ & \\
$P_s$ [day] &  $10.11 \pm 0.02$ & $10.20 \pm 0.01$ &
$10.57 \pm 0.02$ & \\
\hline
\hline
\multicolumn{5}{c}{Differential rotation}\\
\hline
 & Kepler-411b & Kepler-411c & Kepler-411d & all \\
 \hline
$A = \Omega_e$ [rd/d] &  $0.6234\pm0.0013$    &   $0.6230\pm0.0008$  &  $0.619\pm0.010$ & $0.62308\pm0.00022$  \\
$B = \Delta\Omega$ [rd/d]  & $0.0523\pm0.0026$ & $0.0514\pm0.0017$ & $0.043\pm0.019$ & $0.0500\pm0.0006$ \\
$\Delta\Omega/\bar\Omega$  [\%] & $8.8\pm0.4$ & $8.6\pm0.3$ & $7\pm3$ & $8.4\pm0.1$  \\
\hline
\end{tabular}
\tablecomments{{$^{a}$(Wang et al. 2014) $^{b}$ exoplanet.eu $^{c}$This work } }
\end{table*}

\subsection{Starspot modeling}

\begin{figure}[!ht]
    \begin{center}
  \includegraphics[scale=0.5]{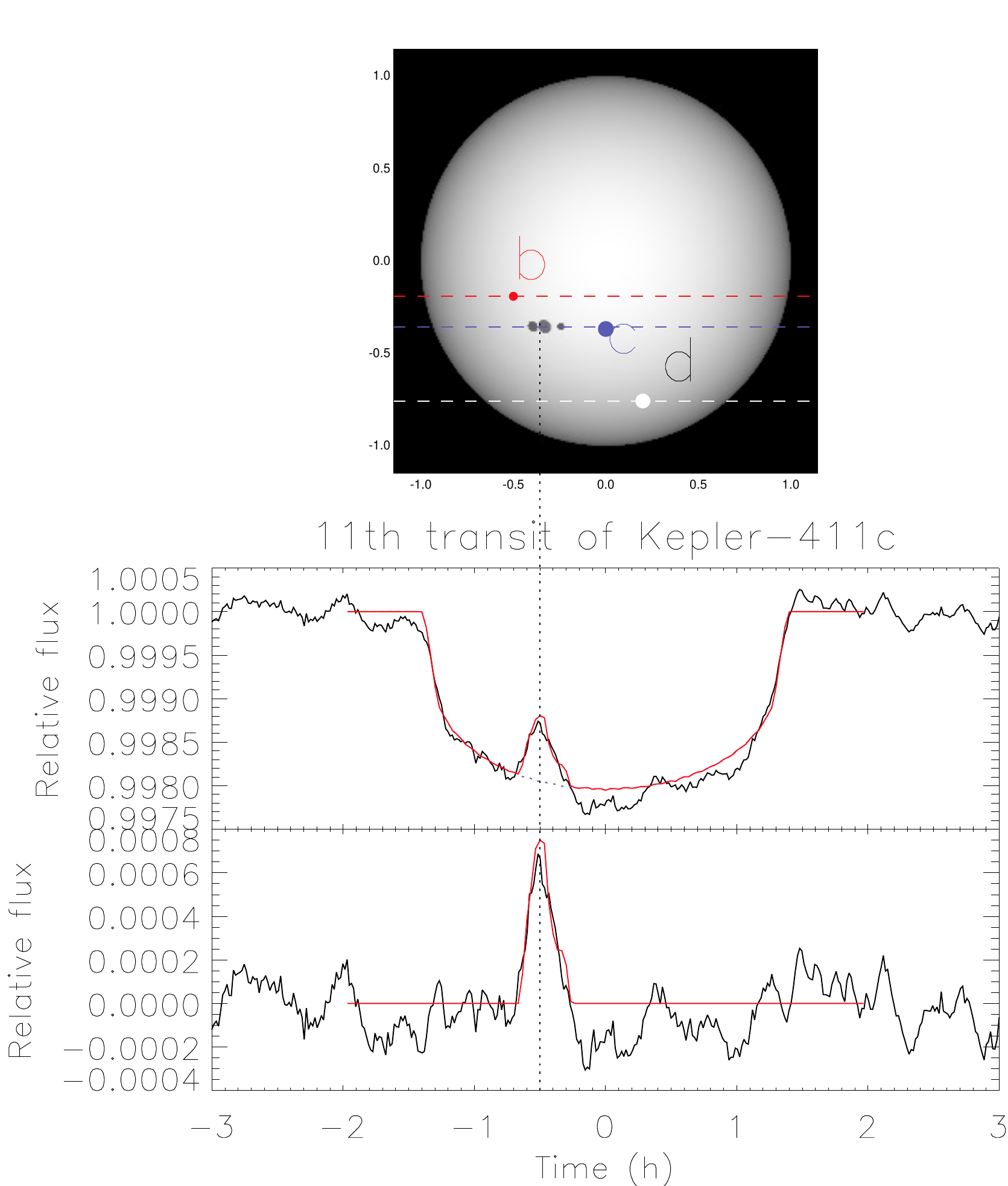}
    \caption{\textbf{Top:} Kepler-411 synthesized stellar disc and its three transiting planets. Planets b, c, and d are represented in red, blue, and white, respectively, along with their transit trajectories (dashed lines). A group of three spots are shown on the surface of the star at the latitude of planet $c$ transit chord. \textbf{Middle panel:} Light curve of Kepler-411c 11th transit with the spots signature, where the red line represents the best fit of the model, and the underlying dotted blue line represents the model of a spotless star.  \textbf{Bottom panel:} Light curve residual, after subtraction of a spotless star model, with the spot model in red. The vertical dotted line connects the location of the spots on the surface of the star with their signatures on the transit light curve, related by Eq.~\ref{eq:lg}.}
    \label{fig:star_planets}
    \end{center}
\end{figure}

In the analysis of Kepler-411 light curves, we isolated the transits of the three planets, b, c, and d. There were 121, 48, and 7 planetary transits detected for Kepler-411b, Kepler-411c, and Kepler-411d exoplanets. In some of these transits, spot signatures could be identified (see Figure~\ref{fig:star_planets}).

Here the starspots were characterized using the spot mapping   model of Silva (2003). The method generates a synthesized two-dimensional image of the star (top panel of Figure~\ref{fig:star_planets}), assuming that the planet is a dark disk with a radius ${R_{p}}/{R_{star}}$, where $R_p$ and $R_{star}$ are the radii of the planet and star, respectively. At each time interval, the position of the planet in its orbit is calculated according to its parameters: angle of inclination \textit{i} and semi-major axis \textit{a}, accounting for the limb darkening of the stellar disk:
\begin{equation}
\frac{I(\mu)}{I_c}=1-u_{1}(1-\mu)-u_{2}(1-\mu)^{2},
\end{equation}
\noindent where $\mu$ is the cosine of the angle between the line-of-sight and the normal to the local surface of the star, and $u_{1}$ and $u_{2}$ are the limb darkening coefficients. $I_c$ is the brightness at the center of the stellar disc.  The Kepler-411 limb darkening coefficients used here are 0.6036 and 0.1164 (Sing 2010). The limb darkened star with spots and the three planets, each in their transiting projected stellar latitudes, are shown in Figure~\ref{fig:star_planets} (planets b, c, and d are represented in red, blue, and white, respectively).

First, for each planet, an averaged light curve of all transits is fit by a spotless star model to determine the parameters: $R_p/R_{star}$, $i$, and $a$. The update planet radius and orbital parameters are listed in Table~\ref{tab:param}.
Next, this spotless model is subtracted from each transit light curve. This is done to highlight spot signatures in the transit, which are seen as peaks in the residuals. Spikes in the residuals with an intensity greater than 3 sigmas are fit. An example of the modeling is shown in Figure~\ref{fig:star_planets} for the 11th transit  lightcurve of Kepler-411c (middle panel) and the residual after subtraction of a spotless star model (bottom panel). The red lines show the best fit from the model. 

\subsection{Starspots and physical parameters}

From the model fit of the peaks detected in the light curve residuals  (bottom panel of Figure~\ref{fig:star_planets}), it is possible to estimate the physical parameters of the Kepler-411 starspots. 

Foreshortenning of the spot when close to the limb is taken into account by the model. Moreover, only spots located within $\pm 70^\circ$ longitudes were considered so as to avoid the steep ingress and egress portions of the light curve. Each starspot is modeled as a circular disc with the following parameters:
\begin{itemize}
\item Radius: size in units of the planetary radius ($R_p$);
\item Temperature: from the intensity with respect to the maximum brightness of the star at disc center, $I_c$;
\item Position: longitude and latitude on the stellar disc. 
\end{itemize}
The latitude, $lat$, and longitude, $lg$, of each spot are given by the projection of the planetary position onto the stellar disc.
Zero longitude is considered as the projection during midtransit, whereas the spot longitude, $lg$, is obtained from the time of the spot peak in the residuals, with $t_s$ given in hours, according to the following equations:
\begin{equation}
lat = \arcsin\left[\frac{a}{R_{star} \cos(i)}\right],
\end{equation}
\begin{equation}
    lg=\arcsin\left[\frac{a\ \cos [90^{\circ}-(360^{\circ}{t_s}/24~ P_{orb})]}{cos(lat)}\right],
     \label{eq:lg}
\end{equation}
\noindent where $R_{star}$ is the radius of the star; and the planet orbital parameters are: $a$  the semi-major axis, $P_{orb}$ the orbital period, and $i$ the orbital inclination angle with respect to the stellar rotation axis. Here we assumed the stellar spin to have null obliquity.

The spot temperature is obtained assuming that both the star and the spot emit radiation like a black body (following Eq. 2 of Silva-Valio et al. 2010), 
where we consider $T_{eff} = 4832$K as the effective temperature of the stellar photosphere and an observing wavelength of $\lambda = 600$nm.

\section{Results} \label{sec:results}
Using the transit method proposed by Silva (2003), we analyzed each transit separately fitting the small variations (peaks) in the flux residuals of the light curve detected during the transits of the three planets. These peaks are interpreted as the signatures of the presence of a spot hidden by the passage of the planet. The detection of the same spot on a later transit allows the determination of the rotation period of the star at the projected transit latitude. The calculated rotation periods at different latitude will then be used to estimate the differential rotation profile of Kepler-411.

\subsection{Modeling of spots Kepler-411}
\begin{figure*}
\centering
\includegraphics[scale=0.35]{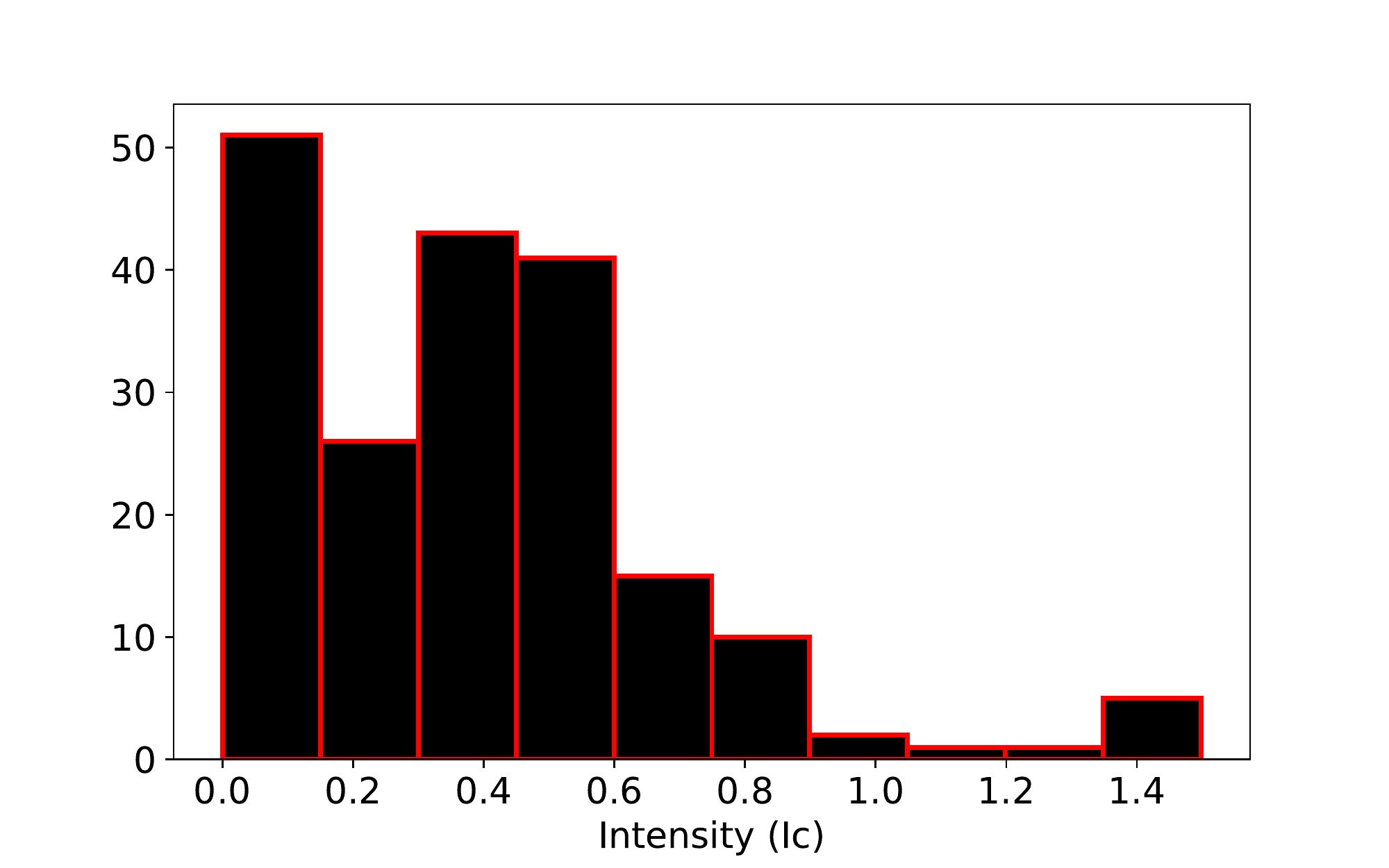}
\includegraphics[scale=0.35]{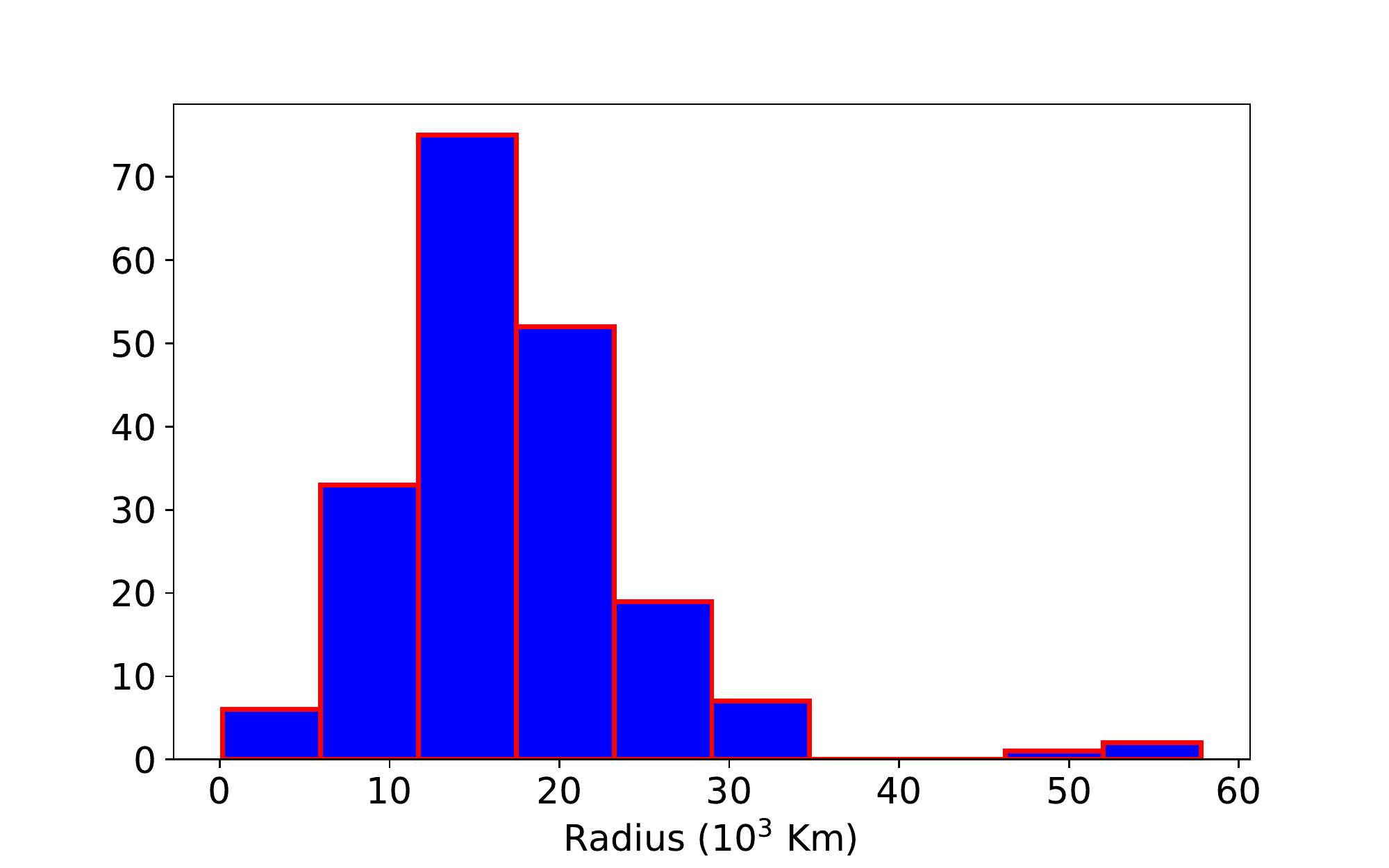}
\includegraphics[scale=0.35]{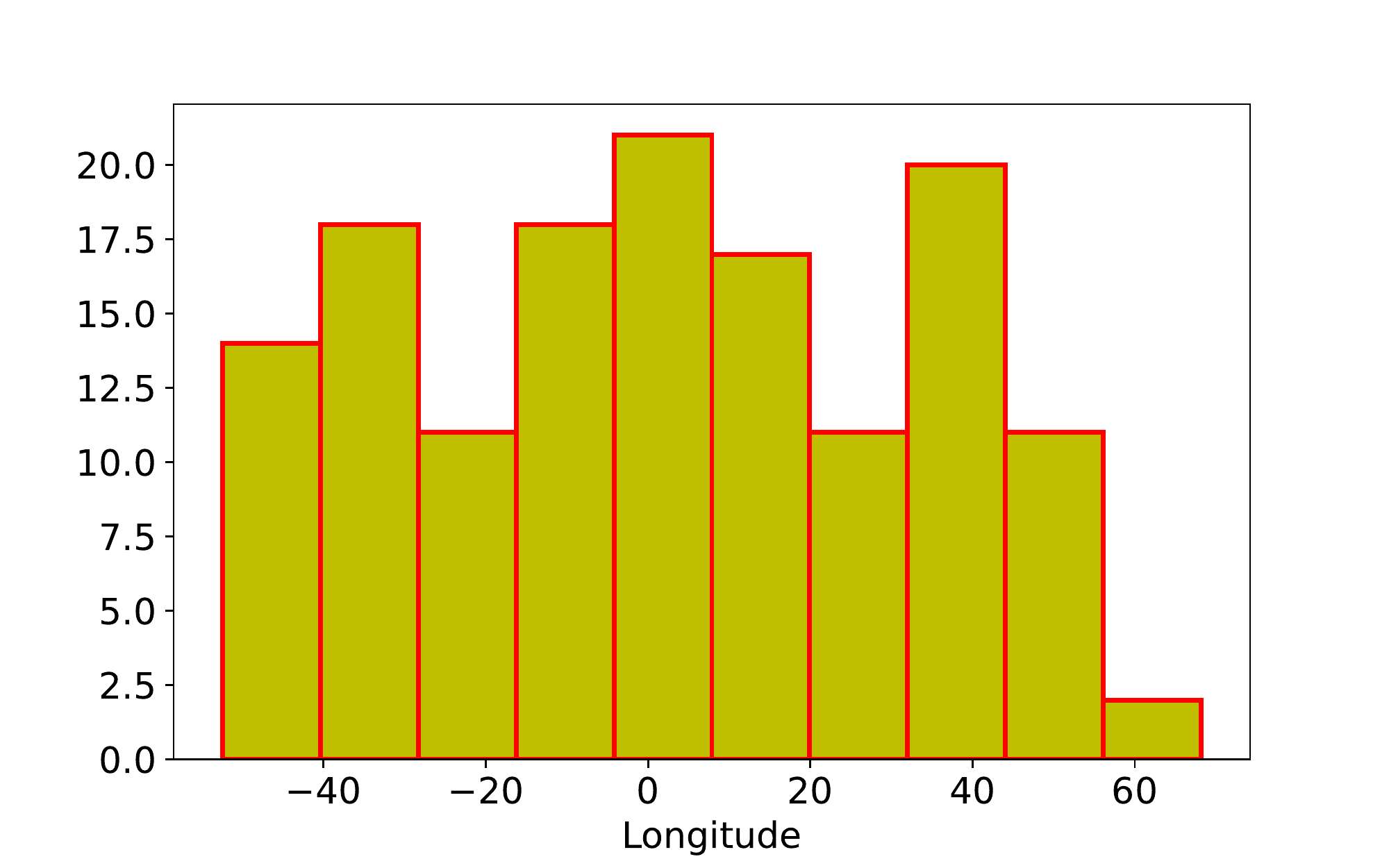}
\includegraphics[scale=0.35]{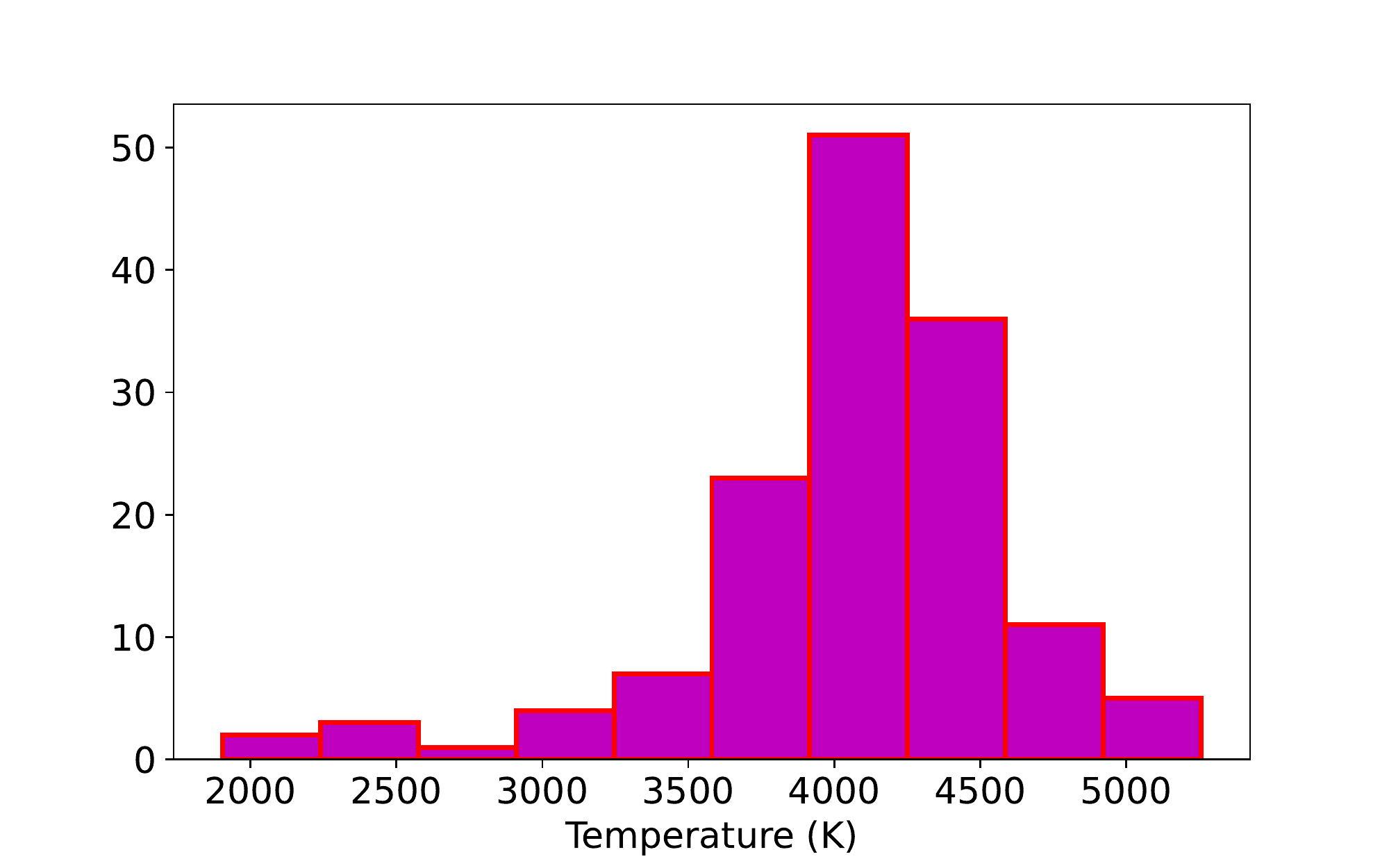}
\caption{Histograms of spot physical parameters: intensity  in units of stellar central intensity, $I_c$ (Black);
radius in km (Blue); longitude on the stellar disc (Yellow); and  temperature in K (Magenta).}
\label{fig:spots}
\end{figure*}

In the analysis of the Kepler-411 light curve, there were 121 (out of 193) light curve transits without data gaps of Kepler-411b, 48 (out of 74) planetary transits of the Kepler-411c, and 7 (out of 10) planetary transits of Kepler-411d. From the transits of the three planets, a total of 198 starspots were identified on the surface of the star, 45 from the passage of Kepler-411b, 143 spots from fitting Kepler-411c transits, and 10 spots from Kepler-411d. 
The distributions of the parameters obtained from the spot modeling are shown in Figure~\ref{fig:spots}, whereas their average physical parameters are listed in Table~\ref{tab:param}. 

\begin{figure*}
    \centering
    \includegraphics[scale=0.32]{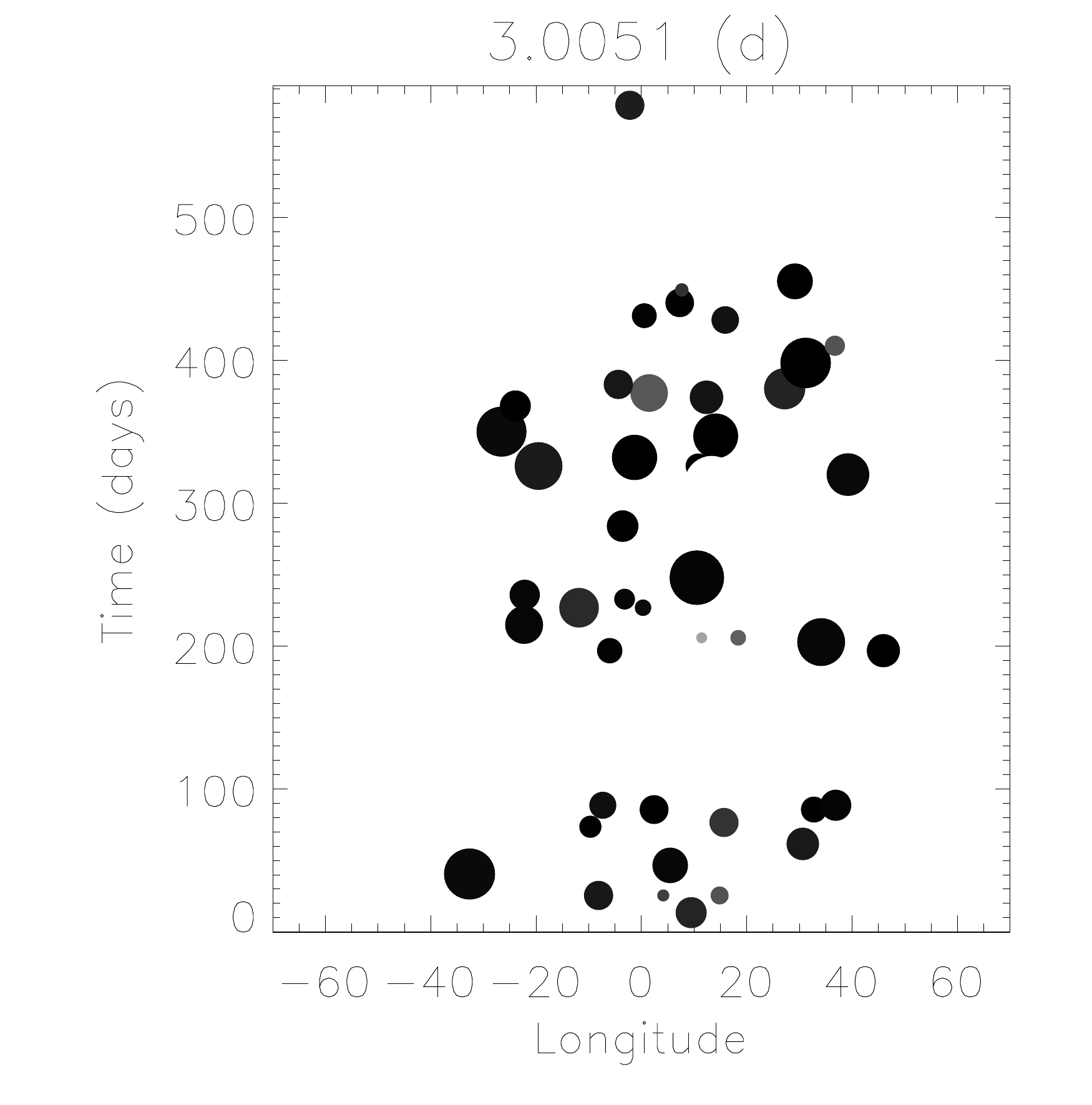}
    \includegraphics[scale=0.32]{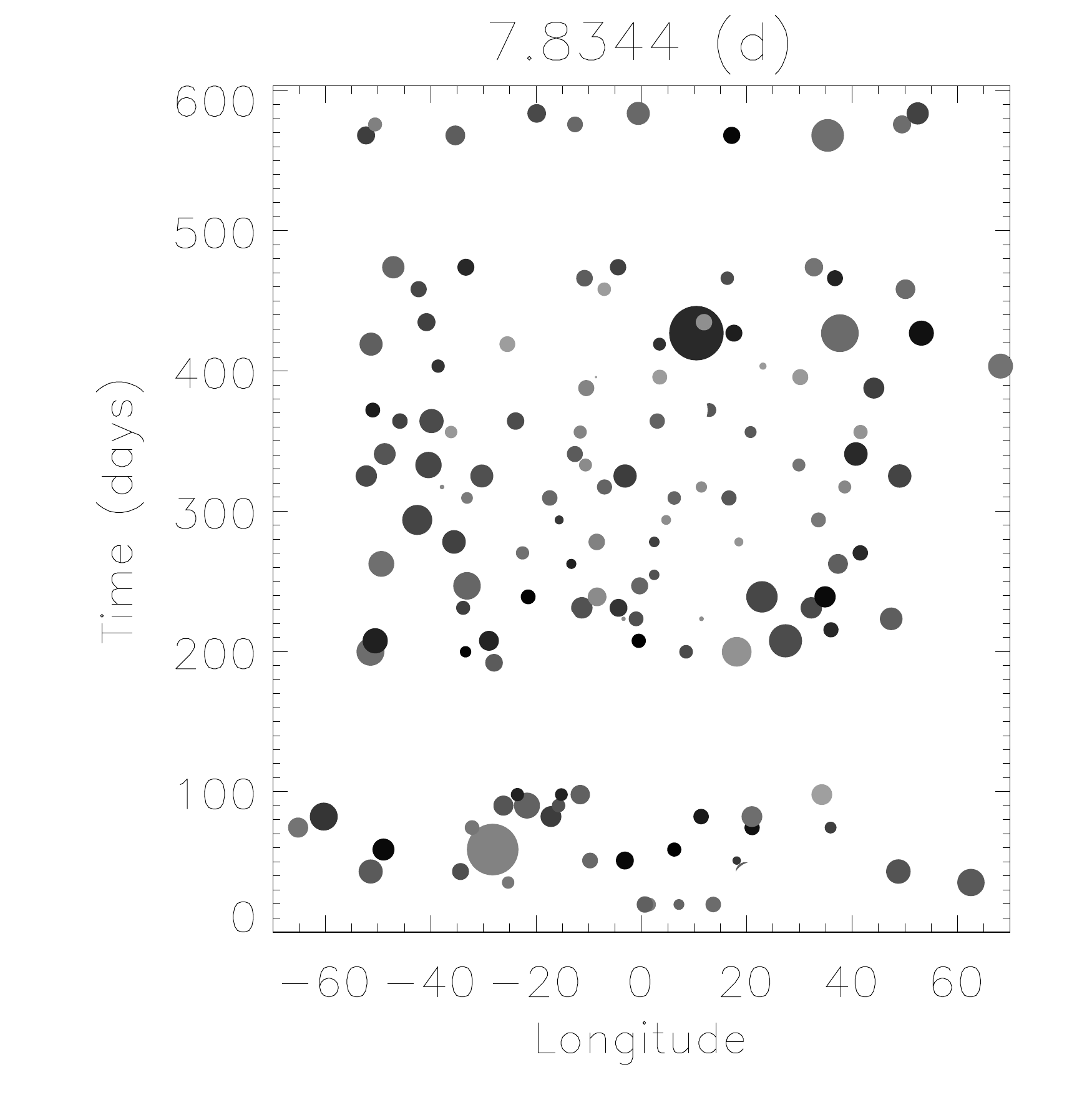}
    \includegraphics[scale=0.32]{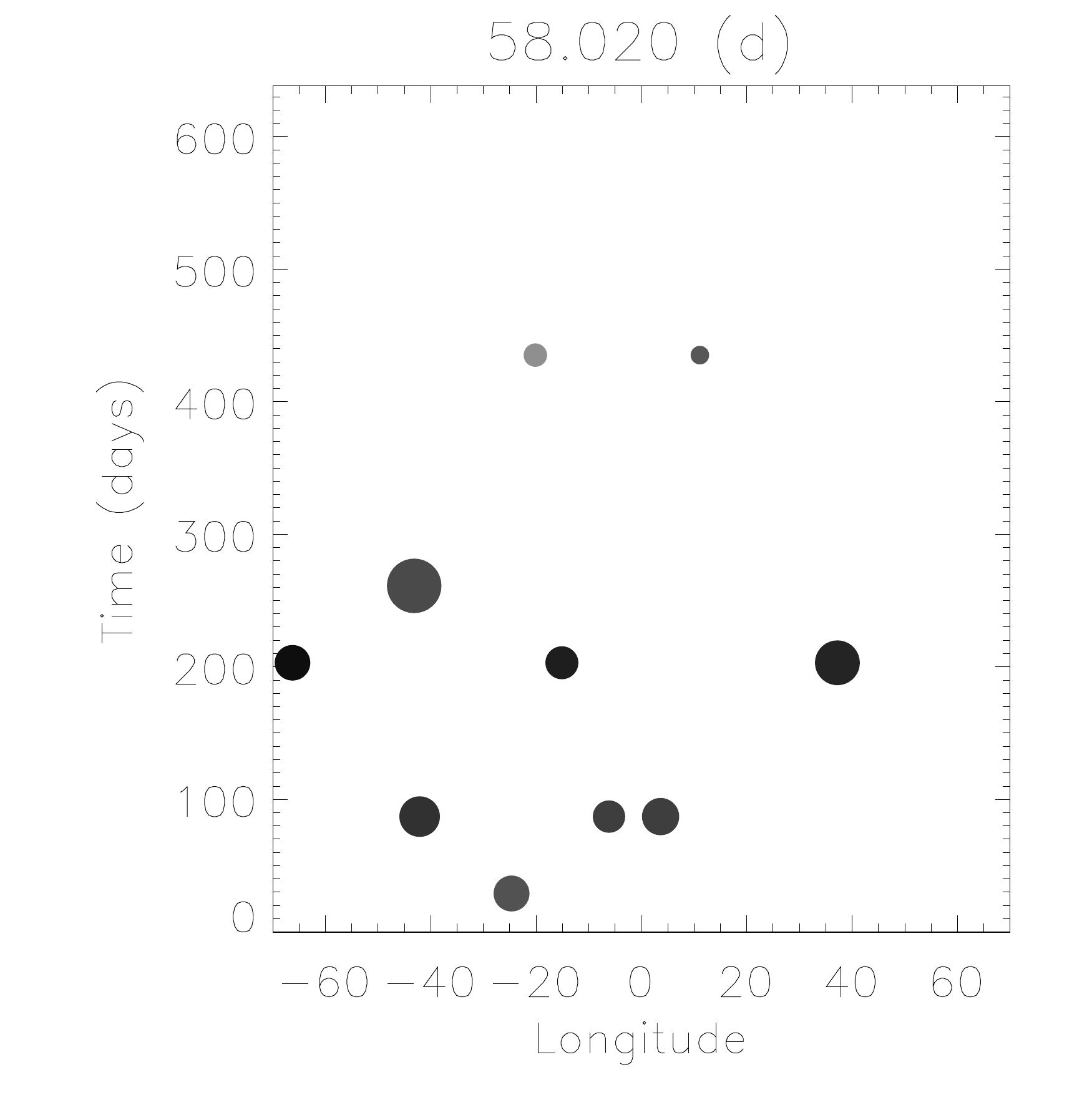}
     \includegraphics[scale=0.32]{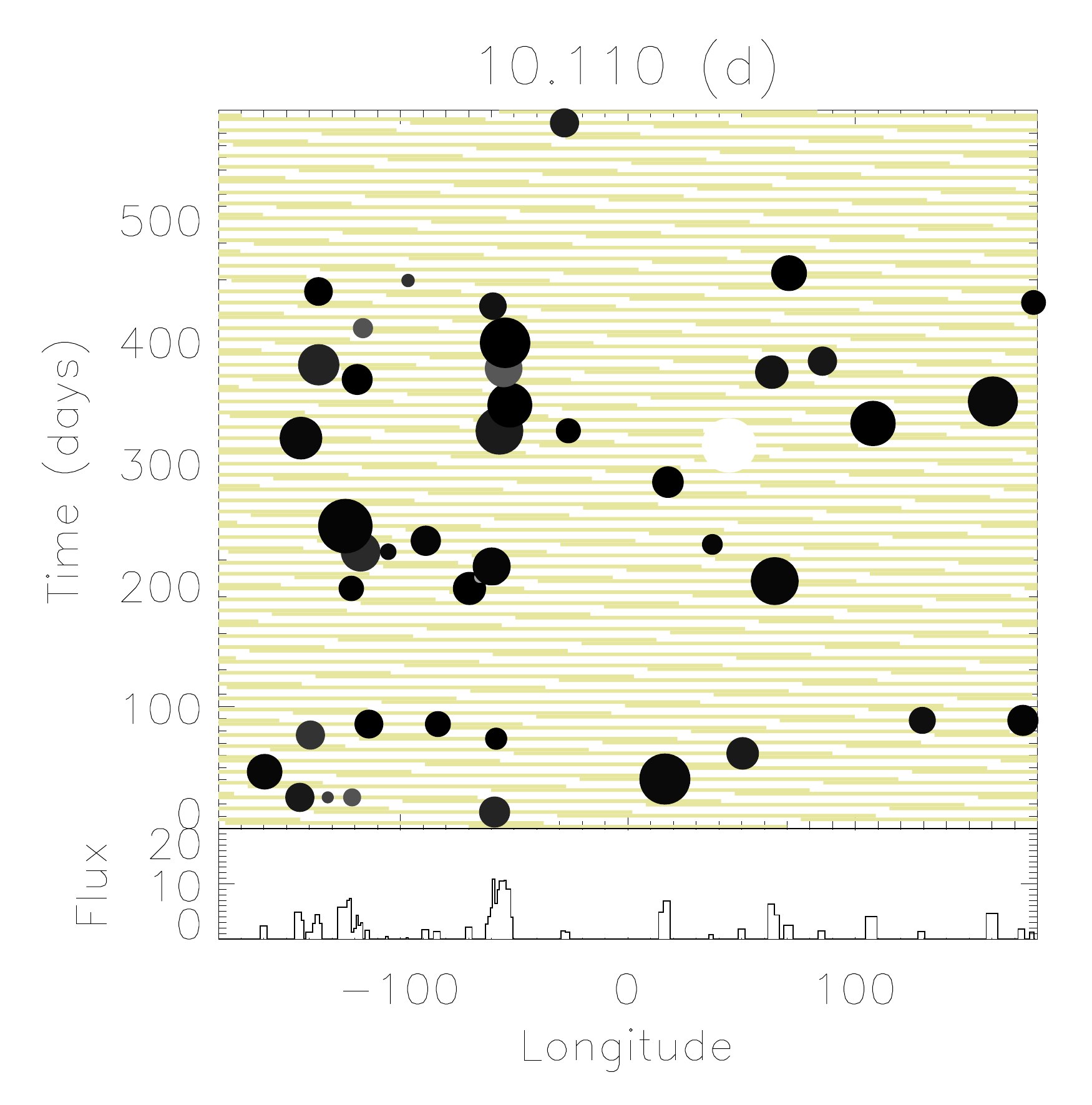}
    \includegraphics[scale=0.32]{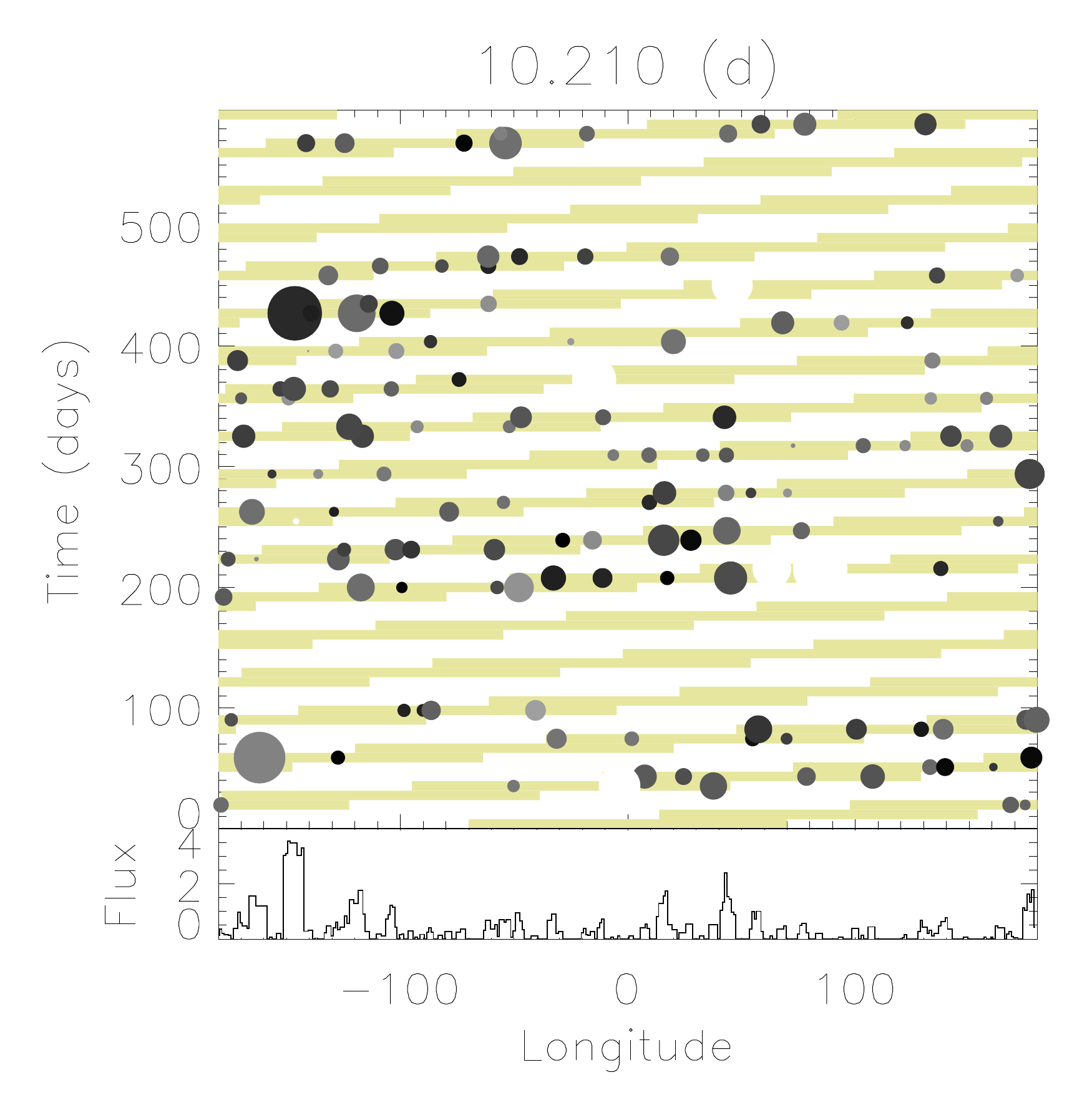}
    \includegraphics[scale=0.32]{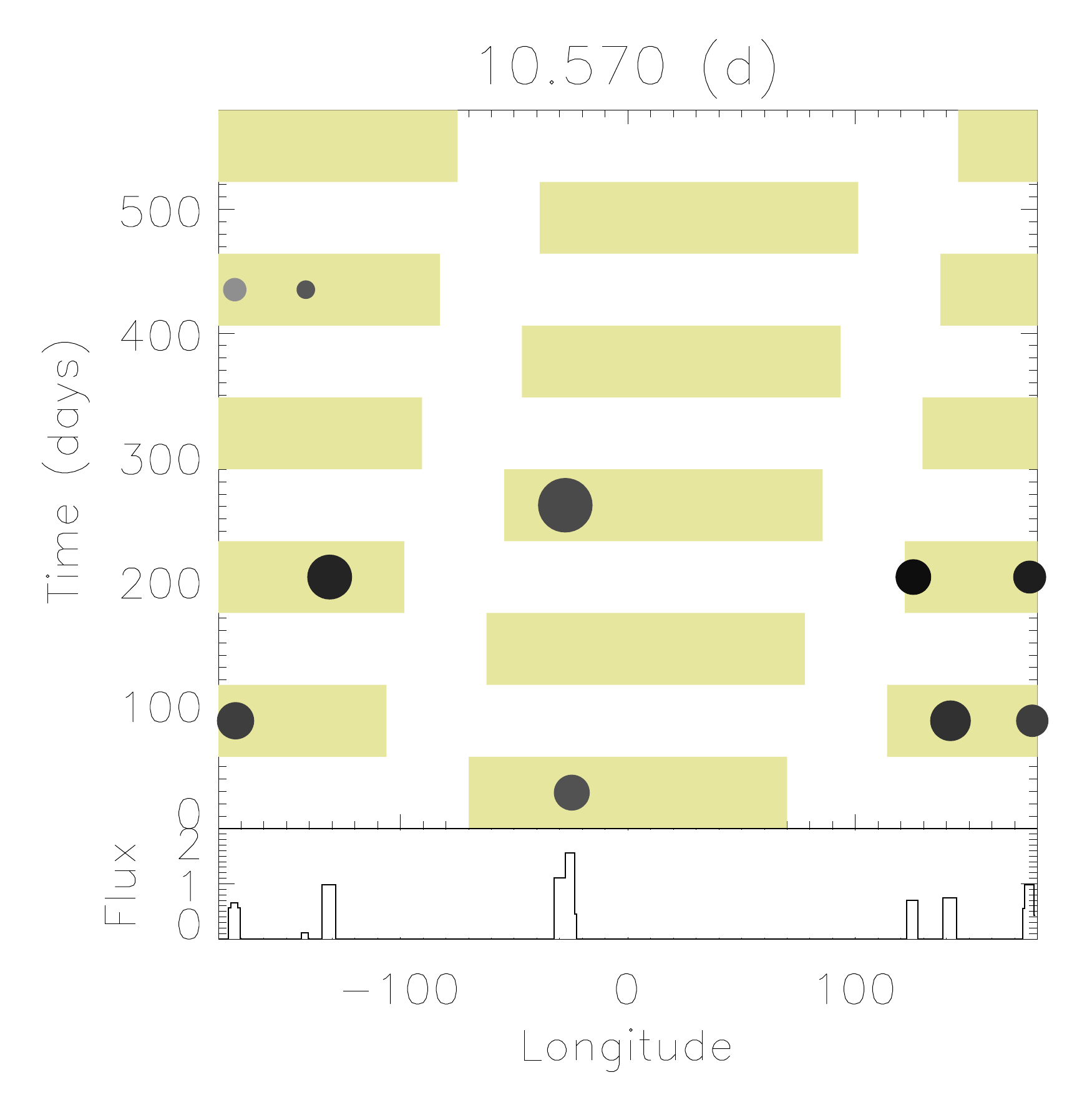}
    \caption{\textbf{Top:} Surface map of Kepler-411 spots at the three latitudes (-11, -21, and 49$^\circ$, from left to right), for topocentric longitudes between -60$^{\circ}$ and 60$^{\circ}$. \textbf{Bottom:} Same as the top panels but in a coordinate system that rotates with the star. The yellow bands depict the visible hemisphere of the star during each transit. Above each map is the period that best describes the rotation period of the star at that latitude. The panels below show the flux deficit due to all the spots at that longitude. }
    \label{fig:map}
\end{figure*}

A one-dimensional map of the stellar surface crossed by each planet in time with the position of the spots viewed from Earth is shown in the top three panels of Figure~\ref{fig:map}. The stellar surface strips occulted by the planet are stacked vertically in time for each planetary transits. Because we only consider modeling of spots within $\pm70^\circ$ due to the steep gradient of the transit light curve during planet ingress or egress, not surprisingly the majority of spots are located within $\pm60^\circ$ longitudes of the stellar disc. The orbital period of each planet is given in the title of the panels. Even though there are many more transits for the closest planet Kepler-411b, because this is the smaller planet, the transit light curves are shallower and the data noise sometimes precludes the identification of spots. On the other hand, because Kepler-411d is farther away, there are fewer transits and not so many spots were detected.

\subsection{Rotation of Kepler-411}\label{sec:rot}

\begin{figure}[ht]
\centering
  \includegraphics[scale=0.44]{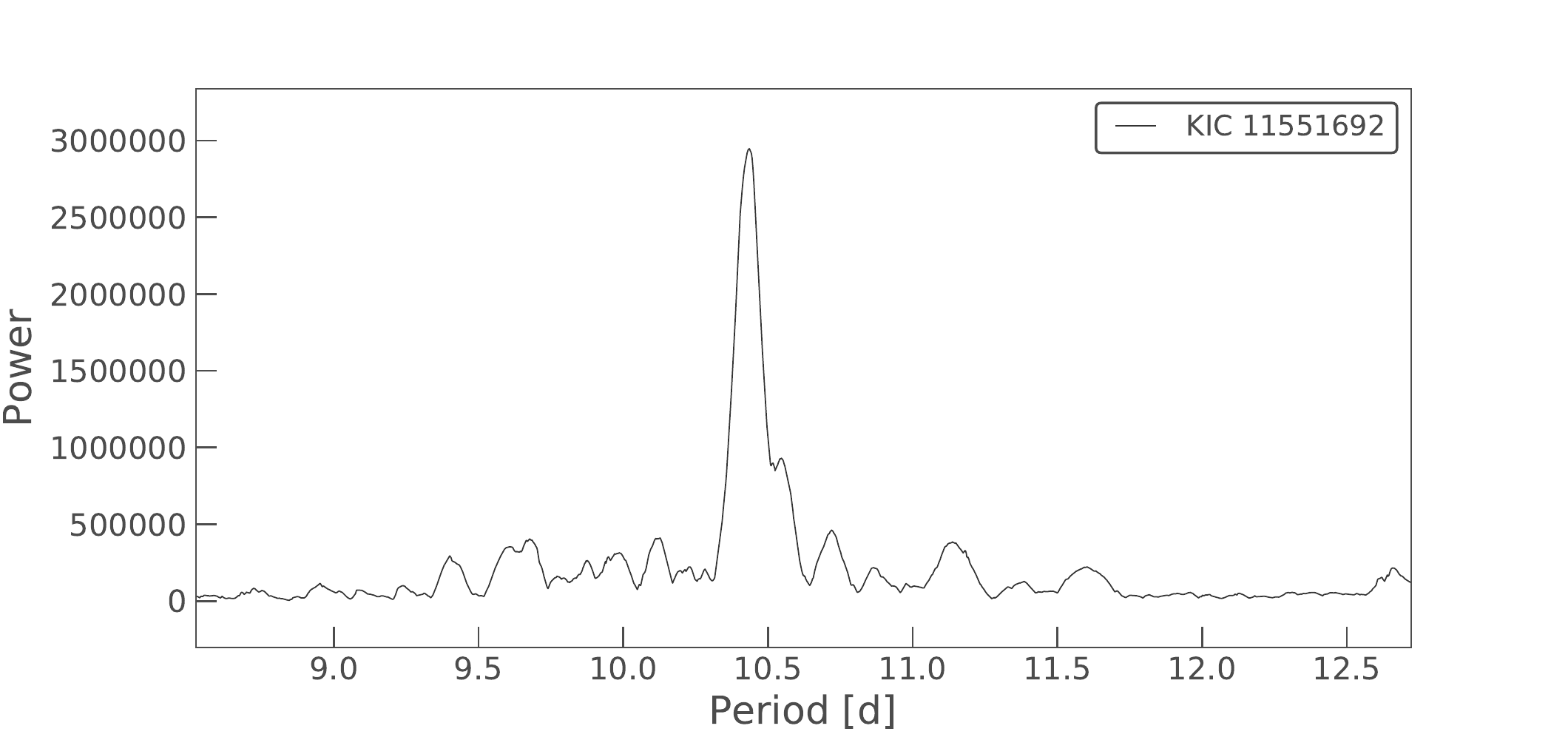}
  \includegraphics[scale=0.5]{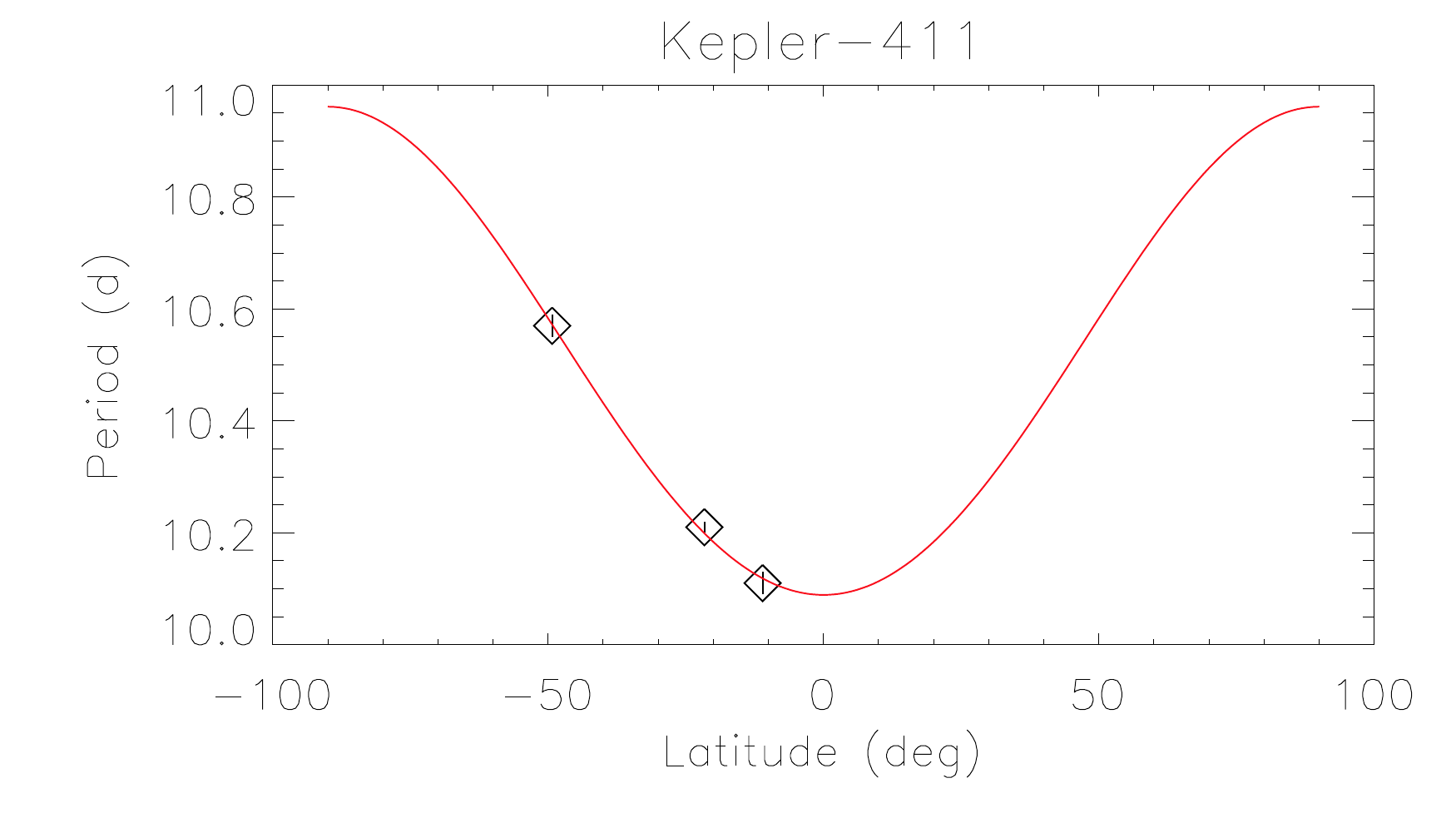}
\caption{\textbf{Top:} Lomb-Scargle periodogram of Kepler-411 light curve, indicating an average rotation period of 10.52 days. \textbf{Bottom:} Differential rotation profile of Kepler-411.} 
\label{fig:rot}
\end{figure}

The rotation of a spotted star creates a modulation in the light curve as spots come in and out of view as the star rotates. 
Through the Lomb-Scargle periodogram it is possible to measure the mean stellar rotation period of  stars.  
The periodogram for Kepler-411 has a strong peak centered at 10.52d, as can be seen in the top panel of Figure~\ref{fig:rot}, which is taken as the average rotation period of the star.

Since Kepler-411 is orbited by 3 transiting planets and each one occults different latitudes of the star (Table~\ref{tab:param}), for the first time it is possible to estimate the rotation period of the star at different latitudes using transit mapping. 
First it is necessary to convert from the topocentric longitude of the spot as viewed from Earth, $lg_{topo}$, to a longitude that rotates with the star, $lg_{rot}$, and therefore varies from 0$^\circ$ to 360$^\circ$:
\begin{equation}
    lg_{rot} = lg_{topo} - 360^\circ \frac{n\ P_{orb}}{P_{s}},
\end{equation}
\noindent where $n$ is the transit number and $P_{s}$ is the rotation period at a given latitude, which is estimated as explained bellow. 

The  spot mapping of each planet transit was analyzed independently. For each latitude band, a 1D map of the stellar surface occulted by the transit (similar to the ones in the bottom row of Figure~\ref{fig:map}) was built for each planet for a given a rotation period, $P_s$. We carried out 100 runs with $P_s$ varying in steps of 0.01d starting at 10d, which covered the average period of 10.52d.
The goal here is to identify in these maps temporal sequences of spots on the same longitude, $lg_{rot}$, of the star (x-axis of the panels), given that some spots will align vertically for the correct $P_s$.  

One can inspect visually the hundreds of maps searching for the aligned spot groups, such as the one in the leftmost bottom panel of Figure~\ref{fig:map} at around $-60^\circ$ longitude from 300--400 days,
or use an automatic method. Here we chose the latter applying it to the flux deficit of spots as a function of longitude obtained from the transit spot maps.
The flux deficit of the spot is defined as $F_{spot} = \Sigma (1-I_{spot})\times R_{spot}^2$, where $I_{spot}$ and $R_{spot}$ are the intensity and radius of each spot, respectively, and the sum is over a given longitude band for all times. Examples of flux deficits are plotted in the bottom part of the lower panels of Figure~\ref{fig:map}. We chose to use flux deficits instead of area or intensity separately due to the degeneracy between these quantities that may result from the modeling. 
Thus, for each value of $P_s$, the total flux deficit in time of the spots at each longitude bin from the map is calculated.

Valio et al. (2017) determined $P_s$ automatically by calculating the auto-correlation function (ACF) of the flux deficit for each period and choosing the one that corresponded to the narrower ACF function as the rotation period of the star at the given latitude.
However, here we determined the period, $P_s$, by maximizing the peaks of flux deficit such that there were fewer peaks but higher in flux. For example, this technique identified the strong peak at $-60^\circ$ of Kepler 411-b (left most panel in the bottom row of Figure~\ref{fig:map}), which results from the clustering of spots at this longitude identified in transits between 300 and 400 days.
The  rotation periods estimated in this way for each transit latitude are listed in Table~\ref{tab:param} and the spotmaps produced are shown in the bottom panels of Figure~\ref{fig:map} .

\subsection{Differential rotation of Kepler-411}

For the stars monitored by the Kepler telescope during long periods of time (months to years), the signatures of the spots during transits can be identified.  This enabled the measurement of the rotational period of the star (Silva-Valio et al. 2010;
Valio et al. 2017; Zaleski et al. 2019) at a certain transit latitude. To estimate the differential
 rotation from a period at the latitude of a single transiting planet,  it is necessary to assume a rotation profile for the star. Here, we consider a solar-like profile:
\begin{equation}
    \Omega = A - B\ \sin^{2}\phi,
    \label{eq:sunprof}
\end{equation}
\noindent where $\phi$ is the stellar latitude and $\Omega = {2\pi}/{P}$ is the rotation rate in rd/d. The rotational shear is given by
$B = \Delta\Omega=\Omega_{eq} - \Omega_{pole}$, where  $A = \Omega_{eq}$ and $\Omega_{pole}$ are the angular rotation rates at the equator and the poles, respectively. To estimate the $A$ and $B$ constants, we need at least two measurements of this profile: (i) the average rotation period, $\bar P$, and (ii) the rotation period at the known transit latitude, $\phi_1$ (Valio 2013).
Once the rotation profile is determined, the differential rotation, $\Delta\Omega$ (rd/d), and the relative differential rotation, ${\Delta\Omega}/{\bar\Omega}$ (\%), where $\bar\Omega = 2\pi/\bar P$, can be calculated for the star.

Here we  estimated the rotation profile of Kepler-411 at a given latitude, as explained in Section~\ref{sec:rot}, using the rotation period obtained from each planet transit spot mapping. The results are listed on Table~\ref{tab:param}. We emphasize that the results were estimated independently for each planet, however, they all agree very well.

Next, since we have measurements of the rotation period at three different latitudes, one for each transiting planet, a profile of the type given by Eq.~\ref{eq:sunprof} was fit to the values of $P_s$ given in Table~\ref{tab:param}. The result of the fit is given in the last column at the bottom part of Table~\ref{tab:param}, and the resulting fit plotted on the bottom panel of Figure~\ref{fig:rot} in red. Again this result agrees very well with that obtained from the transits of a single planet.
In this calculation the average rotation period was not used,  nevertheless estimating the stellar average rotation period from the profile plotted in Figure~\ref{fig:rot}, yields 10.517d, which is very close to the measured 10.52d from the Lomb-Scargle periodogram.

\section{Discussion and conclusions}\label{sec:conclusion}

In this paper, we present a study of stellar activity and rotation for the star Kepler-411, a K2V-type star, with an average rotation period of 10.52 days. This active star hosts three transiting planets, a SuperEarth and two  miniNeptunes with radii of 1.88, 3.27, and 3.31 Earth radii, and periods of 3.0, 7.8, and 58.0 days, respectively. 
Their orbits are such that they transit their host star at three different projected stellar latitudes: -11$^{\circ}$, -21$^{\circ}$, and -49$^{\circ}$.

The occultation of spots seen in the transit light curves of the three planets were characterized using the the model described in Silva (2003). A total of 198 starspots were detected in the transit lightcurves, with an average radius of $17\times 10^3$ km and mean temperatures ranging from 3000--4000 K.
The spots closer to the equator ($-10^\circ$) are the darkest, whereas those at mid latitudes ($-20^\circ$) have smaller contrast with respect to the surrounding photosphere.

Following for example Valio et al. (2017), the rotation period at the projected transit latitude is determined by detecting the same spot on subsequent planetary transits. However, the spot lifetime must be long enough for this to occur, at least longer than the orbital period of the transiting planet, which is usually shorter than the stellar rotation period. 
From analyses of out--of--transit simulated light curves, Basri \& Shah (2020) have shown that spots lifetime can affect  estimates of stellar rotation and differential rotation, especially for short lived spots that last less than 10 rotations. In the case of spot transit mapping, since we rely on detecting the same spot on a later transit, an important timescale here is the orbital period of the planet, $P_{orb}$. It is easier to estimate the stellar rotation at the transit latitude, when $P_{orb}$ is shorter than the stellar rotation period and, of course the spot lifetime, allowing for multiple spot detection within the same stellar rotation (Silva-Valio 2008). This is the case here for planets Kepler-411b and Kepler-411c. Moreover, some of the spots on Kepler-411 seem to last for about 100 days, that is about 10 rotation periods. The study of spots lifetime from transit mapping of Kepler-411 and other stars will be the subject of a forthcoming work.

Independent calculation of the differential rotation profile of the star from the spots identified in each planet transit light curve resulted in similar values for the differential rotation of Kepler-411, assuming a solar like profile. Moreover, the rotation profile obtained from the fit by Eq.~\ref{eq:sunprof} to the rotation periods at the three transiting latitudes was remarkably close to the previous ones. The fit of a solar-like rotation profile yield a differential shear of 0.050 rd/d or a relative differential rotation of 8.4\%, as shown in Figure~\ref{fig:rot} (red curve, bottom panel). This rotational shear is smaller than that of the Sun, $\Delta\Omega_\odot=0.073$ rd/d (Beck 2000).

Previous works have studied the rotation of K stars. Küker \&
Rüdiger (2005) showed that $\Delta\Omega$ depends on the effective temperature of the stars and only weakly on the rotation. Reinhold
et al. (2013) showed that $\Delta\Omega$ ranges from 0.079 rad/d for T$_{eff}$ = 3500K to 0.096 rad/d for T$_{eff}$ = 6000K. In studies of differential rotation of K, G, F, and A stars,  found that differential rotation of stars on the main sequence varies with effective temperature and rotation rate. However, for K and G stars the dependence of  shear on the rotation rate is weak, increasing for F stars, and is strong for A stars.

Fig. 6 of Balona \& Abedigamba (2016) displays $\Delta\Omega$ as a function of $T_{eff}$ for Kepler stars with rotation periods ranging from 6 to 14 days with overlay of models (in red) from Kitchatinov \& Olemskoy (2012) for a star with 10 days rotation period. The value measured here of $\Delta\Omega = 0.05$  for Kepler-411, a star with $T_{eff} = 4832$K is in good agreement with this plot. Applying Eq. 1 of Balona \&
Abedigamba (2016) for the values or $T_{eff}$ and $P_{rot}$ of Kepler-411, we obtain $\Delta\Omega = 0.068$. However, this equation was obtained considering normalized $\Delta\Omega$ with respect to the Sun by applying a factor obtained for solar-type stars.

In a recent study, Xu et al. (2021) report a lower limit for the differential rotation of the Kepler-411. Using Kepler photometry data and spectroscopic data from the Keck I telescope, the authors  analyzed the evolution of spot and differential rotation for the Kepler-411 star using a simple two spot model. As a result they reported a lower limit of $\Delta\Omega = \alpha 2\pi/P_{eq} = 0.065$rd/d. However, Basri \& Shah (2020) have cautioned about using simple two or three spots models for estimating rotation periods and differential rotation.

In this work, from transit spot mapping, we obtained $\Delta\Omega=0.05$ and $\alpha = B/A = 0.080$ (see Table~\ref{tab:param}). These values are lower limits, since they were estimated considering that spots are present on the surface of the star at all latitudes, from 0 to 90$^\circ$. Because of the transit of Kepler-411d, we know that there are spots up until latitude 50$^\circ$. If we consider that the maximum latitudinal range of spots is $75^\circ$, then $\Delta\Omega=0.067$ rd/d and $\alpha = B/A = 0.113$.

We emphasize that the transit mapping method of spots was applied for the first time on a star with multiple transiting planets, thus enabling an accurate estimate of  the differential rotation profile of a star. Moreover, this was also the first time that the spot modeling was performed on a type K star, even though the same technique has been applied to G and M stars also.
Spots characterization and differential rotation are clues to better understand how the dynamo mechanism acts on stars of different masses, thus supporting studies of stars of varying spectral types.

It is important to highlight the precision of the measurement of our work as an important contribution to obtain the differential rotation using the different latitudes of transiting exoplanets.
These findings contribute to our understanding of differential rotation in solar-type stars. More information on stellar differential rotation should help  to establish a greater degree of accuracy in studies of stellar activity, especially the theory of solar dynamo.

\acknowledgments
We are greatful to the anonymous referee for the suggestions that helped improve this work.
The authors acknowledge partial financial support from  FAPESP grant \#2013/10559-5 as well as MackPesquisa funding agency.

\vspace{5mm}
\facilities{Kepler-The Kepler Mission; MAST}


\section{REFERENCES}
\noindent
Appourchaux, T., Michel, E., Auvergne, M., et al. 2008, A\&A, 488, 705
\\
Balona, L. A., \& Abedigamba, O. P. 2016, MNRAS, 461, 497
\\
Basri, G., \& Shah, R. 2020, ApJ, 901, 14
\\ 
Beck, J. G. 2000, SoPh, 191, 47
\\
Berdyugina, S. V. 2005, LRSP, 2, 8
\\ 
Borucki, W. J., Koch, D., Basri, G., et al. 2010, Sci, 327, 977
\\ 
Kitchatinov, L., \& Olemskoy, S. 2012, MNRAS, 423, 3344
\\ 
Korhonen, H., \& Elstner, D. 2011, A\&A, 532, A106
\\ 
Küker, M., \& Rüdiger, G. 2005, AN, 326, 265
\\ 
Lanza, A., Chagas, M. D., \& De Medeiros, J. 2014, A\&A, 564, A50
\\ 
Netto, Y.,\& Valio, A. 2020, A\&A, 635, A78
\\
Pont, F., Sing, D. K., Gibson, N., et al. 2013, MNRAS, 432, 2917
\\ 
Reiners, A., \& Schmitt, J. H. M. M. 2003, A\&A, 412, 813
\\ 
Reinhold, T., Reiners, A., \& Basri, G. 2013, A\&A, 560, A4
\\
Schou, J., Antia, H. M., Basu, S., et al. 1998, ApJ, 505, 390
\\
Silva, A. V. 2003, ApJL, 585, L147
\\ 
Silva-Valio, A. 2008, ApJL, 683, L179
\\ 
Silva-Valio, A., Lanza, A., Alonso, R., \& Barge, P. 2010, A\&A, 510, A25
\\ 
Sing, D. K. 2010, A\&A, 510, A21
\\ 
Sun, L., Ioannidis, P., Gu, S., et al. 2019, A\&A, 624, A15
\\
Takeda, Y. 2020, PASJ, 72, 10
\\
Valio, A. 2013, in ASP Conf. Ser. 472, New Quests in Stellar Astrophysics III: A Panchromatic View of Solar-Like Stars, With and Without Planets, ed.M. Chavez et al. (San Francisco, CA: ASP), 239
\\
Valio, A., Estrela, R., Netto, Y., Bravo, J., \& de Medeiros, J. 2017, ApJ,835, 294
\\
Silva-Valio, A., Lanza, A., Alonso, R., \& Barge, P. 2010, A\&A, 510, A25
\\
Wang, J., Xie, J.-W., Barclay, T., \& Fischer, D. A. 2014, ApJ, 783, 4
\\
Xu, F., Gu, S., \& Ioannidis, P. 2021, MNRAS, 501, 1878
\\
Zaleski, S. M., Valio, A., Carter, B. D., \& Marsden, S. C. 2020, MNRAS,492, 5141
\\ 
Zaleski, S. M., Valio, A., Marsden, S. C., \& Carter, B. D. 2019, MNRAS, 484, 618



\end{document}